\documentstyle[twoside,fleqn,espcrc2]{article}
\newread\epsffilein    
\newif\ifepsffileok    
\newif\ifepsfbbfound   
\newif\ifepsfverbose   
\newdimen\epsfxsize    
\newdimen\epsfysize    
\newdimen\epsftsize    
\newdimen\epsfrsize    
\newdimen\epsftmp      
\newdimen\pspoints     
\def\epsfbox#1{\global\def\epsfllx{72}\global\def\epsflly{72}%
   \global\def\epsfurx{540}\global\def\epsfury{450}%
   \def\lbracket{[}\def\testit{#1}\ifx\testit\lbracket
   \let\next=\epsfgetlitbb\else\let\next=\epsfnormal\fi\next{#1}}%
\def\epsfgetlitbb#1#2 #3 #4 #5]#6{\epsfgrab #2 #3 #4 #5 .\\%
   \epsfsetgraph{#6}}%
\def\epsfnormal#1{\epsfgetbb{#1}\epsfsetgraph{#1}}%
\def\epsfgetbb#1{%
\openin\epsffilein=#1
\ifeof\epsffilein\errmessage{I couldn't open #1, will ignore it}\else
   {\epsffileoktrue \chardef\other=12
    \def\do##1{\catcode`##1=\other}\dospecials \catcode`\ =10
    \loop
       \read\epsffilein to \epsffileline
       \ifeof\epsffilein\epsffileokfalse\else
          \expandafter\epsfaux\epsffileline:. \\%
       \fi
   \ifepsffileok\repeat
   \ifepsfbbfound\else
    \ifepsfverbose\message{No bounding box comment in #1; using defaults}\fi\fi
   }\closein\epsffilein\fi}%
\def\epsfclipstring{}
\def\epsfsetgraph#1{%
   \epsfrsize=\epsfury\pspoints
   \advance\epsfrsize by-\epsflly\pspoints
   \epsftsize=\epsfurx\pspoints
   \advance\epsftsize by-\epsfllx\pspoints
   \epsfxsize\epsfsize\epsftsize\epsfrsize
   \ifnum\epsfxsize=0 \ifnum\epsfysize=0
      \epsfxsize=\epsftsize \epsfysize=\epsfrsize
      \epsfrsize=0pt
     \else\epsftmp=\epsftsize \divide\epsftmp\epsfrsize
       \epsfxsize=\epsfysize \multiply\epsfxsize\epsftmp
       \multiply\epsftmp\epsfrsize \advance\epsftsize-\epsftmp
       \epsftmp=\epsfysize
       \loop \advance\epsftsize\epsftsize \divide\epsftmp 2
       \ifnum\epsftmp>0
          \ifnum\epsftsize<\epsfrsize\else
             \advance\epsftsize-\epsfrsize \advance\epsfxsize\epsftmp \fi
       \repeat
       \epsfrsize=0pt
     \fi
   \else \ifnum\epsfysize=0
     \epsftmp=\epsfrsize \divide\epsftmp\epsftsize
     \epsfysize=\epsfxsize \multiply\epsfysize\epsftmp   
     \multiply\epsftmp\epsftsize \advance\epsfrsize-\epsftmp
     \epsftmp=\epsfxsize
     \loop \advance\epsfrsize\epsfrsize \divide\epsftmp 2
     \ifnum\epsftmp>0
        \ifnum\epsfrsize<\epsftsize\else
           \advance\epsfrsize-\epsftsize \advance\epsfysize\epsftmp \fi
     \repeat
     \epsfrsize=0pt
    \else
     \epsfrsize=\epsfysize
    \fi
   \fi
   \ifepsfverbose\message{#1: width=\the\epsfxsize, height=\the\epsfysize}\fi
   \epsftmp=10\epsfxsize \divide\epsftmp\pspoints
   \vbox to\epsfysize{\vfil\hbox to\epsfxsize{%
      \ifnum\epsfrsize=0\relax
        \includegraphics{#1}%
      \else
        \epsfrsize=10\epsfysize \divide\epsfrsize\pspoints
        \includegraphics{#1}%
      \fi
      \hfil}}%
\global\epsfxsize=0pt\global\epsfysize=0pt}%
{\catcode`\%=12 \global\let\epsfpercent=
\long\def\epsfaux#1#2:#3\\{\ifx#1\epsfpercent
   \def\testit{#2}\ifx\testit\epsfbblit
      \epsfgrab #3 . . . \\%
      \epsffileokfalse
      \global\epsfbbfoundtrue
   \fi\else\ifx#1\par\else\epsffileokfalse\fi\fi}%
\def\epsfempty{}%
\def\epsfgrab #1 #2 #3 #4 #5\\{%
\global\def\epsfllx{#1}\ifx\epsfllx\epsfempty
      \epsfgrab #2 #3 #4 #5 .\\\else
   \global\def\epsflly{#2}%
   \global\def\epsfurx{#3}\global\def\epsfury{#4}\fi}%
\def\epsfsize#1#2{\epsfxsize}
   \newcommand{\cqq}{ {\cal Q}^{\dagger}{\cal Q} }
\newcommand{\ra}{\rangle}
\newcommand{\la}{\langle}
\newcommand {\dm}{ \Delta {\cal M}^{-1} }
\newcommand{\be}{\begin{equation} }
\newcommand{\beq}{\begin{eqnarray} }
\newcommand{\eeq}{\end{eqnarray} }
\newcommand{\ee}{\end{equation} }
\newcommand{\eps}{\epsilon}

\newcommand{\AmS}{{\protect\the\textfont2
  A\kern-.1667em\lower.5ex\hbox{M}\kern-.125emS}}

\hyphenation{author another created financial paper re-commend-ed}

\title{Hubbard Model and L\"{u}scher fermions }

\author{P. Sawicki\address{Institute of Physics, 
        Jagellonian University, \\ 
        Reymonta 4, Cracow, Poland}
       }

\begin{document}

\begin{abstract}
We discuss numerical complexity of the L\"{u}scher
algorithm applied to the Hubbard Model. In particular
we present comparison to a certain  algorithm,
based on direct computation of the fermionic determinant.

\end{abstract}

\maketitle

\section{Introduction}

Recently a new type of algorithms for numerical
simulations of models with dynamical fermions
has been proposed by M.~L\"uscher \cite{lu94}.
The idea is based on the approximation
of a fermionic determinant by the path integral
over some number of bosonic fields, which allows ones to
elegantly avoid the main difficulty coming from 
the nonlocality of the fermionic determinant. 
The complexity of a sweep of the L\"uscher algorithm 
is proportional to the volume of the system and
in this respect it wins in comparison with the standard
hybrid Monte Carlo \cite{ken}. Of course, what is
interesting in practice is a comparison of 
the computer cost needed to produce independent 
configurations which is the product of the 
sweep cost and the autocorrelation time. 
Therefore the crucial question is how 
the autocorrelation time depends on the volume of the 
system. Contrary to the sweep time the autocorrelation
time can strongly depend on the external parameters.
Therefore it is particularly interesting to test the
dependence of the autocorrelation time on the
various interesting physical parameters.
Recent results obtained in QCD are rather promising showing
that the computational cost needed to obtain independent 
configurations is comparable to the standard HMC 
algorithms \cite{alex}. 

In this letter we present new results of numerical
simulations of the Hubbard Model using
the L\"uscher method and compare them to standard methods
in context of the algorithmical complexity.
This is a continuation of our contribution 
to the proceedings Lattice 95 where we presented
preliminary results \cite{saw95}. In particular,
we noted there possible improvements 
which reduce autocorrelation times.
Here, we study the errors introduced by 
the bosonic approximation and compare dynamical properties of the
multiboson algorithm with the simple exact algorithm. 

\section{L\"uscher approximation }

We consider the Hubbard Model defined by the Hamiltonian
\begin{eqnarray}
{ \cal H}=-K\sum_{ \la ij \ra \sigma} a_{i\sigma}^{\dagger} a_{j\sigma}
-{U\over 2}\sum_i (n_{i\uparrow} - n_{i\downarrow})^2   
\end{eqnarray}
where  $a_{i\sigma}$ are fermionic operators acting on electrons 
on site $i$ of the lattice. The sum runs over nearest neighbors
(each symmetric pair $ \la i,j \ra $ is counted twice). 
Physical parameters entering the model are: K - hopping parameter,
U - the strength of the effective Coulomb interaction, 
and the inverse temperature $\beta$.
The interaction term of the Hamiltonian 
describes the system with the half-filled band. One find the
following representation of the partition function for
the model \cite{saw94}~:
\begin{eqnarray}
  \lefteqn{ Z \simeq \int [dA d\phi]   \exp{ \left( -\int    A^2(x)/2 \;
     d^3x \right)  }   } \nonumber \\
   \lefteqn{\exp{\left( -\int  \sum_{k=1}^{n}
\phi^{\dagger}_k 
\left[   (\cqq-\alpha_k)^2+\beta_{k}^{2}  \right] \phi_k  d^3 x  \right) 
    }    }.  \nonumber  \\ \label{zet} 
\end{eqnarray}
where ${\cal Q^{\dagger} \cal Q}\equiv
 {\cal M ^{\dagger} \cal M }/\lambda_{max}$ and $\lambda_{max}$ 
 denotes the largest eigenvalue of the matrix 
 ${\cal M}^{\dagger}{\cal M}$. The matrix ${\cal M}$ is 
 the fermion matrix entering the Euclidean formulation
\begin{eqnarray}
\Psi^{\ast} {\cal M} \Psi = {K\beta\over N_t}
\sum_{\la ij \ra t} \Psi^{\ast}_{it}\Psi_{jt} + 
\sum_{it}\Psi^{\ast}_{it}(\Psi_{it}-\Psi_{it-1})   \nonumber \\ +
\sum_{it}\Psi^{\ast}_{it}\Psi_{it}
\left( \exp{ \left[
\sqrt{U\beta\over N_t} A_{it} - \frac{U\beta}{N_t} \right]} -1 
\right) . \nonumber
\end{eqnarray}
$A(x)$ is the Hubbard-Stratanovich continuous field, and $\phi_k(x)$ are
the auxiliary bosonic fields. The particular form of the 
polynomial $P_n (x)$ used in the approximation of the function
$1/x$, is built from the Chebyshev polynomials and gives a rapid 
uniform convergence to the function $1/x$ in the interval $( \eps ,1)$. 
Its roots are introduced already in (\ref{zet}) and their
real and imaginary parts are denoted by $\alpha_k + i \beta_k, k=1,...,n$ 
respectively. The relative error of the approximation defined
as $|R(x)|=|x P_n(x) -1|$ is exponentially bounded
\be
   |R(x)| < 2  \left( \frac{ 1- \sqrt{ \epsilon}}{1+ \sqrt{ \epsilon}}
              \right)^{n+1}.
\ee
One sees that to decrease $\eps$ one has to increase $n$. 
The most economic choice of $\eps$ corresponds to the largest possible
value which still gives a good approximation of the determinant. 
To make this statement precise 
we introduce following the quantity \cite{for95} as 
a measure of error (or goodness)
\be
\delta = | y^{1/V} -1 |
\ee
where
\be
y= \det{ (Q^\dagger Q)  P_n(Q^{\dagger} Q) } .
\ee
The power $1/V$, where $V$ is naturally adjusted to the 
dimension of the matrix ${\cal M}$.
Figure 1 shows $\delta$ as a function of $\eps$ for a different
numbers of fields measured on a one typical configuration. The optimal
value of $\eps$ lies close to the smallest eigenvalue of the matrix
$Q^\dagger Q \; \;  (\lambda_{min}=0.032) $. Such a behavior is 
consistent with the previous findings in the QCD case.  

\section{Exact algorithm}

To compare the computational cost of the
multiboson algorithm with the cost of the more standard approaches,
we implemented a simple exact algorithm based on the updating 
scheme allowing for computing changes in the matrix ${\cal M}^{-1}$
while making trial changes in the matrix ${\cal M}$. In order to update the
field $A$ located on the site $i$ one must calculate the ratio of the
fermion determinants. Since the only $A$ dependent elements of the matrix
${\cal M} $ lie on the diagonal we consider the following change in the 
matrix ${\cal M}$,
\be
{\cal M'} = (I + \Delta) {\cal M},
\ee
where $\Delta$ is matrix with one nonzero element 
$\Delta_{i j}= \delta_{i j} d$. Then 
\be
\frac{ \det {{\cal M'} } } { \det {{\cal M}} }= \det{ (I + {\cal M}^{-1} 
\Delta) }
= 1 + {\cal M}^{-1}_{i i} d
\ee
is determined completely by elements of the matrix ${\cal M}^{-1}$. 
Once the trial change has been accepted the updated matrix 
${\cal M}^{-1}$ can be evaluated from the formula 
\be
 {\cal M}^{' -1}= {\cal M}^{-1} - \frac{{\cal M}^{-1} \dm}
 {1 +d {\cal M}^{-1}_{i i} }.
\ee

\begin{figure}
\epsfxsize=8cm
\epsfbox{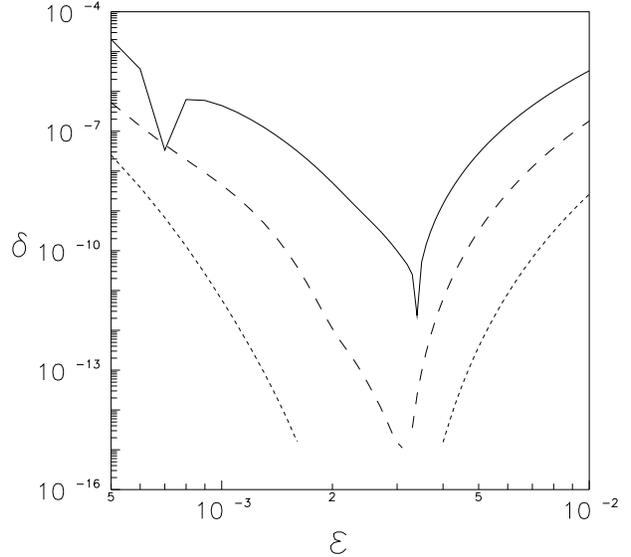}
\caption{Errors of the L\"uscher approximation measured on a single
typical configuration. The quantity $\delta$ defined in text is
shown as a function of $\eps$. The solid, dashed and dotted
line is for number of fields 50, 70 and 100 respectively }
\end{figure}

\begin{figure}
\epsfxsize=8cm \epsfbox{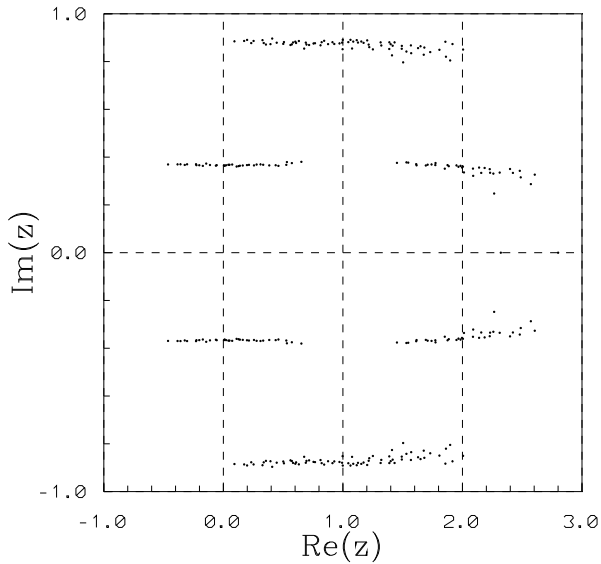}
\caption{Eigenvalues of the matrix ${\cal M}$ on the complex plane }
\end{figure}

This process is very economical from the computational point of view since
one update of $M^{-1}$ requires $O(V^2)$ operations  comparing with
$O(V^3)$ operations needed  to evaluate the determinant with the brute
force method.  Also the today's leading algorithm 
for simulations of the Hubbard Model is essentially 
based on it \cite{sugar}. However, contrary 
to the standard one our particular  formulation of path integrals does not
require  multiplication badly conditioned matrices. Thus we believe that 
this is free of the numerical instabilities. Indeed, 
we performed thousands sweeps
at $\beta$ up to 8 without accumulating numerical errors. An
additional cost comes from working with larger matrices.

\section{Final remarks}

As we have previously reported the strong autocorrelation between generated
configurations is a serious obstacle which limits the available range of
parameters to $U \leq 2$ and $\beta =1$ for the multiboson algorithm. In
contrast the program based on exact evaluation of the fermionic
determinant has no such restrictions and works equally  
good at $U=1$ and $U=4$.

The comparison of CPU time needed for both algorithms varies accordingly to
the lattice sizes. On small lattices exact algorithm is substantially faster.
For example for a lattice $6^2 \ 8$ in our experiments on 
the workstation HP735/125 it takes 20 minutes which should
be compared with 10 hours needed for our implementation of
the L\"uscher algorithm to produce results of the same quality. Of
course for larger lattices the computational cost grows much more faster
for the exact algorithm. In the region of weak coupling we were able to
could simulate with the L\"uscher algorithm lattices with the size 
up to $16^2 \ 8$. 

The polynomial approximation of the function $1/x$ can be extended to the
complex plane \cite{monty}. Thus one try to approximate directly the inverse
of the matrix ${\cal M}$. This would reduce the  condition numbers for matrices
entering the problem and would result in the simpler final action. This
modification of the original L\"uscher idea gave 
very promising results \cite{galli}. 
However, in contrast to the QCD case the matrix ${\cal M}$  for the
Hubbard Model has eigenvalues with the positive and negative real parts
as can for example be seen in Figure 2. Because one cannot 
extend the domain of the convergence beyond the singular point 
(0,0), this is, unfortunately, impossible to adopt this
modification to the Hubbard Model.

\begin{center}
{\bf Acknowledgments }
\end{center}

I would like to thank J.~Wosiek for for many stimulating discussions
and Z.~Burda for reading the manuscript.
Work supported in part by the KBN grant PB 2P03B19609.
Numerical simulations were
partly performed in the ACK Cyfronet-Krak\'ow. Computational grant: 
KBN/UJ/055/94.

\end{document}